\documentclass{aa}

\usepackage{graphicx} 
\usepackage[varg]{txfonts} 
\usepackage{hyperref}

\newcommand{\kms}{\text{km}\,\text{s}^{-1}}
\newcommand{\kmsmpc}{\text{km}\,\text{s}^{-1}\,\text{Mpc}^{-1}}
\newcommand{\diff}{\text{d}} 
\newcommand{\sub}[1]{_\text{#1}} 
\newcommand{\supp}[1]{^\text{#1}}
\newcommand{\thetas}{\theta\sub s} 
\newcommand{\Dd}{D\sub d}
\newcommand{\Ds}{D\sub s} 
\newcommand{\Dds}{D\sub{ds}}
\newcommand{\Deff}{D\sub{eff}} 
\newcommand{\tildeDeff}{\tilde D\sub{eff}} 
\newcommand{\Deffcorr}{\Deff'}
\newcommand{\deffcorr}{\deff'} 
\newcommand{\deff}{d\sub{eff}}
\newcommand{\propmot}{\sub{p.m.}}
\newcommand{\vecnablatheta}{\vec\nabla_{\!\theta}}
\newcommand{\msun}{M_\sun}

\newcommand{\Omegam}{\Omega\sub M} 
\newcommand{\Omegaw}{\Omega_w}

\newcommand{\ph}[1]{\phantom{#1}}

\begin{document}

   \title{Cosmology with gravitationally lensed repeating fast radio
     bursts}

   \author{O. Wucknitz\inst{1}
     \and L. G. Spitler\inst{1} \and
     U.-L. Pen\inst2\fnmsep\inst3\fnmsep\inst4\fnmsep\inst5\fnmsep\inst1
   }

   \institute{Max-Planck-Institut f\"ur Radioastronomie, Auf dem
     H\"ugel 69, 53121 Bonn,
     Germany\\ \email{wucknitz@mpifr-bonn.mpg.de} \and Canadian
     Institute for Theoretical Astrophysics, University of Toronto, 60
     St.~George Street, Toronto, Ontario M5S 3H8, Canada \and Dunlap
     Institute for Astronomy and Astrophysics, University of Toronto,
     50 St.~George Street, Toronto, Ontario M5S 3H4, Canada \and
     Canadian Institute for Advanced Research, CIFAR Program in
     Gravitation and Cosmology, 661 University Ave, Toronto, Ontario
     M5G 1Z8, Canada \and Perimeter Institute for Theoretical Physics,
     31 Caroline Street North, Waterloo, Ontario, N2L 2Y5, Canada }

   \date{Received 24 April 2020 / Accepted 16 October 2020}

   \abstract{High-precision cosmological probes have revealed a small
     but significant tension between the parameters measured with
     different techniques, among which there is one based on time
     delays in gravitational lenses. We discuss a new way of using
     time delays for cosmology, taking advantage of the extreme
     precision expected for lensed fast radio bursts (FRBs), which are
     short flashes of radio emission originating at cosmological
     distances. With coherent methods, the achievable precision is
     sufficient for measuring how time delays change over the months
     and years, which can also be interpreted as differential
     redshifts between the images. It turns out that uncertainties
     arising from the unknown mass distribution of gravitational
     lenses can be eliminated by combining time delays with their time
     derivatives. Other effects, most importantly relative proper
     motions, can be measured accurately and disentangled from the
     cosmological effects. With a mock sample of simulated lenses, we
     show that it may be possible to attain strong constraints on
     cosmological parameters. Finally, the lensed images can be used
     as galactic interferometer to resolve structures and motions of
     the burst sources with incredibly high resolution and help reveal their
     physical nature, which is currently unknown. }

   \keywords{gravitational lensing: strong -- cosmology -- distance
     scale}

   \maketitle

\section{Introduction}

The gravitational lens effect \citep{refsdal64a} can deflect light so
strongly that it produces multiple images of a single
source. \citet{refsdal64b} argued that time delays (light travel time
differences between images) can be used to determine distances and,
thus, the Hubble constant ($H_0$), at a time when it was not even clear
that the effect would ever be seen in observations. For very distant systems,
\citet{refsdal66} showed that time delays can also be used to test
cosmological theories.

The practical applications of this brilliantly simple idea turned out to
be quite difficult. Even after the discovery of the first
gravitationally lensed active galactic nucleus (AGN) by
\citet{walsh79}, it took many years before the time delay between the
two images was agreed upon with sufficient accuracy. The main reasons for
this difficulty come from the typically slow intrinsic variations of AGN as compared to the lensed supernovae that were originally proposed.

An even more fundamental difficulty is the influence of the a priori
unknown mass distribution of the lens on the derived
results. Parameterised mass models can often be fitted to the observed
image configuration, flux ratios, and relative image distortions. In
the best case, extended sources with rich substructures provide a
wealth of constraints for realistic multi-parameter mass
distributions. However, there are fundamental degeneracies between the
lensing mass distribution and the unknown intrinsic source structure.

Best-known is the mass-sheet degeneracy \citep{falco85,gorenstein88},
according to which scaling a given mass distribution while adding a
homogeneous mass sheet can leave the image configuration unchanged,
provided that the source structure and position is scaled
accordingly. This transformation also scales the time delay and thus
the derived Hubble constant. More realistic, but very similar in
effect, are changes of the radial mass profile of the lens.  A more
general degeneracy has been described by \citet{schneider14}.

Additional measurements of the velocity dispersion of lensing galaxies
can partly break these degeneracies, but at the cost of introducing
additional complex astrophysics into the problem. Nevertheless,
very competitive results have been achieved so far. \citet{wong19}
describe results from a joint analysis of six gravitational
lenses. Their result for the Hubble constant is precise to 2.4~\%, but
disagrees with cosmic microwave background (CMB) results far beyond
the formal uncertainties \citep{planck18}. \citet{millon19} discuss
systematic uncertainties in the lensing analysis. \citet{kochanek20}
presents a more pessimistic view and argues that accuracies below
10~\% are hardly possible, regardless of the formal precision.

\vspace{0.5ex} 
Even though the tension between determinations of the Hubble constant
with different methods is well below 10~\% at present, it is still highly
significant. These differences may result from a limited understanding
of the systematics involved or, for instance, a certain behaviour exhibited by dark
energy that is different than that assumed based on simple models. Either way, there is a
need for additional clean methods with preferably less (or, at least, different) systematics. In this work, we argue that a novel application of gravitational lensing is a very promising option. The unique properties
of fast radio bursts (FRBs) are essential for this approach, which diverges from related ideas presented in the
literature. For example, \citet{li18} and \citet{liu19} propose using the high
accuracy of time delays measured from gravitationally lensed FRBs
together with classical mass modelling. They argue that the absence of
a bright AGN core in FRB host galaxies facilitates the use of optical
substructures as lens modelling constraints.

This essential new observable -- namely, time delay changes over time -- has been
discussed by \citet{piattella17} for rather special
lenses. \citet{zitrin18} explicitly investigate the fact that time derivatives
of time delays (as expected from cosmic expansion) may be measurable
with lensed FRBs. They also discuss the notion that the relative proper motion of the
source has a strong effect that is difficult to distinguish from
cosmological expansion.

Here, we argue that with coherent time-delay measurements, we can achieve levels of accuracy that are much
better than the burst duration  and not only does this allow us to disentangle cosmological effects from the proper motion,
but we can even eliminate the unknown mass distributions of the
lensing galaxies as the main source of potential systematic errors.

This manuscript is structured as follows. In
Section~\ref{sec:frbs}, we describe the sources (FRBs) that can be used
to determine time delays at a level of precision that even allows us
to measure how they evolve over time. In Section~\ref{sec:theory}, we
describe how these new observables can be used to avoid the effects of the unknown mass distribution almost entirely. It turns out that
relative transverse motion has a stronger effect than cosmology. We discuss how this effect can be removed, with the additional result of
measuring the proper motion with unprecedented accuracy.
In Section~\ref{sec:several lenses}, we discuss the level to which we may
be able to determine combinations of cosmological parameters based on
realistic measurements. We find that with an ensemble of lens systems,
we can potentially obtain competitive constraints on a number of
parameters.

At the level of precision that is required and achievable, many
additional small effects have to be considered. These possible caveats
are the subject of Section~\ref{sec:caveats}. Similar to the proper
motion, these are also interesting in themselves.
The use of gravitationally lensed images as arms of a galaxy-size
interferometer is discussed in Section~\ref{sec:gravlens
  interferometer}. We can potentially resolve structures of a few
kilometres at cosmological distances. At the same time, the required
time-delay precision is only achievable for sources that actually have
structures on these scales, which is only true for fast radio bursts.
A  discussion and summary is presented in Sec.~\ref{sec:discussion}.

\begin{table*}[t]
  \caption{Typical order-of-magnitude values referred to in the text}
  \label{tab:numbers}
  \centering
  \begin{tabular}{lcl}
    \hline\hline Quantity & Value & Comment \\ \hline time delay &
    $10^6$ s & $\sim 12$ days \\ time delay uncertainty & $10^{-6}$ s
    & conservative \\ Hubble constant & $2\times 10^{-18}$ s$^{-1}$ &
    $H_0 = 70~\kmsmpc$ \\ time between bursts & $10^8$ s & $\sim 3$
    years \\ relative time delay change & $2\times 10^{-10}$ & between
    bursts, due to Hubble expansion\\ lensing redshift & $2\times
    10^{-12}$ & due to Hubble expansion\\ differential redshift
    uncertainty & $10^{-14}$ & from time-derivative of time
    delays\\ image separation & 1 \text{arcsec} \\ transverse speed &
    300 $\kms$ \\ proper motion & $6\times
    10^{-8}~\text{arcsec}\,\text{yr}^{-1}$ & transverse speed at
    1\,Gpc \\ motion-induced redshift & $5\times 10^{-9}$
    \\ Galactocentric acceleration $/c$ &
    $8\times10^{-19}~\text{s}^{-1}$ & $250~\kms$ at 8.5~kpc,
    \citet{macmillan19} \\ acceleration of lens $/c$ & $2.6\times
    10^{-20}~\text{s}^{-1}$ & LMC acceleration for Milky Way \\ \hline
  \end{tabular}
\end{table*}

\vspace{0.5ex} 
\section{Fast radio bursts (FRBs)}
\label{sec:frbs}

A new type of radio sources was discovered by \citet{lorimer07}, which
are bright radio bursts with durations of less than a few milliseconds.
Similarly to pulsars, their signals are dispersed; they arrive later at
lower frequencies. In FRBs, this dispersion is higher than expected
from the integrated electron column density within the Milky Way,
which is evidence for an extragalactic origin at cosmological
distances. The first of these objects were detected only once, which
makes an accurate localisation difficult. The discovery of the first
repeating FRB by \citet{spitler16} changed the game
completely. Observing the roughly known location with the Karl
G.\ Jansky Very Large Array (VLA) allowed \citet{chatterjee17} to
localise it to sub-arcsecond accuracy. This was sufficient to identify
a host galaxy and thus determine the redshift $z=0.1927$ of the host and
FRB \citep{tendulkar17}. This comprised the final proof that (at least) this
FRB originates at a cosmological distance. \citet{marcote17}
determined the position to milli-arcsec (mas) accuracy with Very Long
Baseline Interferometry (VLBI).

The first direct localisation at discovery of another FRB source was
achieved with the Australian Square Kilometre Array Pathfinder (ASKAP)
by \citet{bannister19}, which led to the identification of a host
galaxy at $z=0.3214$. The same instrument is now finding new FRBs
regularly. An even higher yield is produced by the Canadian HI
Intensity Mapping Experiment \citep[CHIME,][]{amiri18}, which does now
find several new FRBs per day. CHIME also discovered the second
repeating FRB \citep{amiri19}, so far without an accurate
localisation. The first VLBI-localisation of another repeater found by
CHIME was presented by \citet{marcote20}.

Given that these instruments, even though they are considered wide-field, can
only observe a small fraction of the sky, the number of FRBs
that are detectable by a true all-sky survey with sensitivities comparable to
current facilities must at least be hundreds, if not thousands per
day.  At the same time, radio astronomical technology is making rapid
progress, which makes a full-sky FRB monitoring feasible in the
foreseeable future.

Currently, the redshift distribution can only be estimated, which makes
it difficult to predict the fraction that is gravitationally
lensed. The Cosmic Lens All Sky Survey \citep[CLASS,][]{browne03}
found that one out of 700 of their AGN source population is strongly
lensed, with multiple images on arcsecond scales. \citet{marlow00}
studied the redshift distribution of the CLASS parent source
population and found a wide range with a mean of $z=1.2$ and a root mean square (rms)
scatter of 0.95. The known source redshifts from \citet{browne03} are
0.96, 1.013, 1.28, 1.34, 1.39, 2.62, 3.214,  and 3.62.  Even though the FRB
population may be systematically closer to us, which reduces the
chance of gravitational lensing, the future discovery of lensed FRBs
is a realistic possibility. Some fraction of them will be repeating,
which will allow us to measure how time delays evolve over time and to apply
the approach presented in this work.
We note that we have to distinguish between the concepts of bursting
FRB sources, individual bursts from FRB sources, and lensed `echoes'
(or images) of individual bursts.

With lensed bursts of millisecond durations, it is obvious that time
delays can be measured with at least this level of accuracy. Typical time
delays are on the order of days to months, and for order-of-magnitude
estimates, we assume $10^6$~s, that is, approximately 12~days; see
Table~\ref{tab:numbers} for the numbers that we use for our
estimates. The fractional accuracy is then $10^{-9}$, a huge leap
forward compared to the few-percent level that is (at the utmost)
achievable with lensed AGN. \citet{millon20} present a table of known
and new time delays of quasars lensed by single
galaxies. Uncertainties are mostly above one day. A particularly good
time delay for B0218+357 is known from gamma-ray observations
\citep{cheung14} and radio monitoring \citep{biggs18}. The results are
consistent with each other and have uncertainties of 0.16 and 0.2
days, respectively. Significant improvement beyond this is not
expected for lensed AGN.

As \citet{li18} argue, the uncertainty of the time delay measurements
for lensed FRBs is entirely negligible in the total error
budget. Because the lensed host galaxies of FRBs are generally not
expected to harbour a strong AGN, their structure in optical images may make lens
modelling more accurate as well, so that a level of 1\% might be
reachable with ten lensed FRBs \citep{li18}.
We do have some concerns that even if this approach might certainly reduce some
uncertainties, the fundamental mass-model degeneracies persist and  could
lead to significant systematic errors even in a joint analysis of many
lenses. For this reason, we propose an alternative route for cosmology
with lensed FRBs.

We need an even higher accuracy of the time delays to follow this
route. Fortunately, this is made possible by lensed FRBs. If we can observe
not only the intensity as function of time, but the full
electromagnetic wave field from a source, and if we have two coherent
copies of the wave field from the same burst, then we can determine
the time delay (group delay) between them with an uncertainty given by
the reciprocal bandwidth. With good signal-to-noise ratio (S/N), the
accuracy can be improved even more. Modern radio receivers have
bandwidths of hundreds of MHz up to a few GHz, so that uncertainties
smaller than a nanosecond are certainly achievable. In principle, we
may even use the (absolute) phase delay, with uncertainties given by
the reciprocal observing frequency.\footnote{When using this absolute
  phase delay, a frequency-independent phase shift depending on the
  type of image has to be taken into account. In principle, this could
  even be used to determine the types of image.} In reality, practical
problems like clock drifts as well as ionospheric and atmospheric delays have
to be considered.  We show below that as result of
time-varying time delays, different images will have different
redshifts. For short bursts, this effect is so small that it generally
does not have to be taken into account in the correlation analysis,
but can easily be included when needed. What has to be included, on
the other hand, are differences of dispersion in the interstellar
medium of the lensing galaxy that produce frequency-dependent delay
differences. These follow a $\lambda^2$ law and can be corrected with
a coherent de-dispersion.

Achieving nanosecond accuracy over time ranges of weeks and months is
not entirely unrealistic, but we base our arguments on a much more
conservative assumption of uncertainty, namely at the level of one microsecond
(Tab.~\ref{tab:numbers}), which corresponds to a fractional
uncertainty of the typical time delay on the order of $10^{-12}$. We note
that this coherent way of measuring time delays does not rely on
short bursts, or even on the presence of intrinsic intensity
variations, because it uses the wave field itself. It is, however,
essential that the images are mutually coherent, which is not the case
for any sources other than FRBs, as explained in
Section~\ref{sec:gravlens interferometer}. Of course finding the time
delay between the waves is made much easier when we have a good estimate
from the intensity correlation.

\section{Theory}
\label{sec:theory}

\subsection{Homogeneously expanding Universe}

Our general idea is very simple: As result of the expansion of the
Universe, time delays should also roughly increase at this rate,
$\diff\Delta t/\diff t\sim H_0\,\Delta t$, which over about three
years (Tab.~\ref{tab:numbers}) corresponds to a change of 0.2~ms or a
fractional change of $2\times 10^{-10}$, which is easily measurable
with an accuracy of better than one percent.  In the following, we
derive how this time-delay increase is related to cosmological
parameters and observables.

For the moment, we assume that the entire Universe, including the
gravitational lens, expands uniformly with a geometry according to the
following metric:\footnote{We use a slightly sloppy notation in which
  the functions $R(t)$, $R(\eta)$ and $R(z)$ are written with the same
  letter.}
\begin{align}
  \label{eq:metric 1}
  \diff s^2 &= c^2\,\diff t^2 - R^2(t) \, \diff L^2 ,\\ &= R^2(\eta)
  \left( \diff \eta^2 - \diff L^2 \right) ,\\ \diff \eta &=
  \frac{c\,\diff t}{R(\eta)} ,\\ \diff L^2 &= \sum_{i,j=1}^3
  \gamma_{ij}\, \diff x_i\,\diff x_j.
  \label{eq:metric 2}
\end{align}
Here $t$ denotes cosmic time (also proper time of comoving observers),
$R(t)$ is the scale factor, and $\diff L$ is the comoving length
interval. The comoving spatial coordinates are $x_j$ and the spatial
metric, $\gamma_{ij},$ does not depend on time.

Under this condition, light rays follow spatial geodesics with a
time-dependence defined by the speed of light. Light rays connecting
the comoving source and the observer at different times follow the same
spatial path and are shifted by a constant interval in the conformal
time coordinate, $\eta$, which provides the important relation between
scale factor and redshift,
\begin{equation}
  1+z = \frac{R_0}{R(z)} .
\end{equation}    
In the case of a gravitational lens, the time delay will not vary over
time if measured in terms of $\eta$. Because the measured time delay,
$\Delta t,$ is a scaled version of $\Delta\eta$, its logarithmic
time-derivative is a direct measurement of the current Hubble
constant:
\begin{equation}
\left.  \frac{1}{\Delta t}\frac{\diff \Delta t}{\diff t}\right|_0 =
\left. \frac{\diff \ln\Delta t}{\diff t}\right|_0 =
\left. \frac{\diff\ln R}{\diff t}\right|_0 = H_0(t)
\label{eq:dot dt simple}
.\end{equation}
In this unrealistic situation, the time delay increases exactly in proportion with
the expanding Universe. We could thus measure the Hubble constant
directly, without mass model uncertainties and even without knowing
image positions or redshifts. In reality the gravitational lens itself
does not expand with the Hubble flow and we have to derive the effect
in detail as shown in the following.

\subsection{Lensing theory}

With the image position $\vec\theta$, the true source position
$\vec\thetas$ and the apparent deflection angle $\vec\alpha$, the lens
equation reads
\begin{equation}
  \vec\thetas = \vec\theta - \vec\alpha (\vec\theta) ,
  \label{eq:lenseq}
\end{equation}
where all angles are small two-dimensional vectors in the tangential
plane relative to some reference axis.

Time delays can be derived using the methods described by
\citet{cooke75}. For this, we do not need to assume a certain
cosmological model as long as we parameterise it by angular size
distances between observer and lens $\Dd$, between observer and source
$\Ds$, and between lens and source $\Dds$. Here, we use the subscripts
d and s for deflector and source, respectively, and suppressed 0 for
the current epoch (observer).  We assume that the global geometry of
the Universe still follows the homogeneous expansion according to
Eqs.~\eqref{eq:metric 1}--\eqref{eq:metric 2}. Later in this paper, we include
the effect of radial motion into these parameters.

An angular size distance, $D\sub{ab}$, is defined as the ratio between a
transverse offset at b and the corresponding angle measured at a,
where both sides are considered at the cosmic time at which the signal
passes.  Because the spatial geodesics do not change in comoving
coordinates and because the length is measured at b, this distance
scales with $R\sub b$ over time, as long as a and b are fixed in
comoving coordinates.  We later relax the assumption that light
rays follow geodesics of the average global spacetime and allow for
systematic overdensities (or underdensitites) near the light path.

The time delay measured by an observer consists of a geometric component
and the Shapiro (or potential) component,\footnote{With $+\text{const,}$ we
  denote an additive term that does not depend on $\vec\theta$ and
  therefore cancels in time delays between images. This constant can
  be different in different equations.}
\begin{align}
  c\,\Delta t_0 &= \Deffcorr \left[
    \frac{(\vec\theta-\vec\thetas)^2}{2} - \psi(\vec\theta) +
    \text{const} \right] ,
  \label{eq:timdel}
  \\ \Deff &= \frac{\Dd\Ds}{\Dds} , \quad \Deffcorr = (1+z\sub
  d)\,\Deff .
\end{align}
Later on, we absorb the $\thetas^2/2$ term into the constant.

The Shapiro delay is described via the lensing potential $\psi$, which
is related to the delay, $\tau\sub d$, as measured in the lens frame via
\begin{equation}
  c\,\tau\sub d = -\Deff \, \psi .
  \label{eq:psi Shapiro}
\end{equation}
The lens Equation \eqref{eq:lenseq} results from Fermat's principle
with $\vec\alpha=\vecnablatheta\,\psi$.

\subsection{Lensing-induced redshifts}

The time delay will slowly change over time in a way that is more
complicated than in the relatively the naive equation in Eq.~\eqref{eq:dot dt simple}. Details are
presented in the following.

The time derivative of the time delay can also be interpreted as
additional lensing-induced redshift $z(\vec\theta)$, in the sense that
the time lag between subsequent bursts in image $\vec\theta$ scales
with $1+z(\vec\theta)$. This scaling only has a meaning as relative
redshift between images, $z(\vec\theta_1)-z(\vec\theta_2)$, similar to
the time delay itself, which only has a unique definition as delay
between images.

The lag between bursts (in contrast to the delay between lensed images
of the same burst) cannot be measured coherently and, therefore, has a
much lower accuracy on the order of a millisecond. Because we measure the
redshift as the difference of time derivatives of time delays, which are
measured coherently, the achievable accuracy is actually given by the
ratio of the coherent timing uncertainty and the burst separation. If
we use Latin letters as indices for images and Arabic numbers for the
bursts (dropping the subscript 0 in $t_0$), the relation can be
written as:
\begin{multline}
  z\sub A - z\sub B \approx \frac{1+z\sub A}{1+z\sub B}-1 =
  \frac{(t\sub{A2}-t\sub{A1})-(t\sub{B2}-t\sub{B1})}{t\sub{B2}-t\sub{B1}}
  =
  \\ \frac{(t\sub{A2}-t\sub{B2})-(t\sub{A1}-t\sub{B1})}{t\sub{B2}-t\sub{B1}}
  \approx \frac{\Delta t\sub{AB}|_2-\Delta t\sub{AB}|_1}{t_2-t_1}
  ,
\end{multline}
which is correct to the first order in $z$ and thus sufficient for all
practical purposes. For real measurements, we can use as many bursts as
available and fit them jointly.

Based on the values in Table~\ref{tab:numbers}, the measurement accuracy in
$z$ is on the order of $10^{-14}$ even under our conservative
assumptions.  This is better than for the relative change of the time
delay because the relevant baseline is the time between bursts, which
can be many years, and which increases without limit if the FRB is
repeating persistently.  It is this extreme precision that makes
gravitationally lensed FRBs such a promising tool. Because we want to
measure a tiny effect (time-delay increase due to the expanding
Universe) with extreme precision, we have to consider many other
effects that may influence the results. In the analysis below we find
no show-stoppers, but some of the additional effects (most importantly
the proper motion) are interesting even in themselves.

\subsection{Distance parameters in a Robertson-Walker Universe}

For most of the final analysis, we assume that the general geometry
(with the exception of the lens) is expanding homogeneously according
to Eqs.~\eqref{eq:metric 1} and \eqref{eq:metric 2}. This means that
the angular size distance, $D\sub{ab}$, can be written as a comoving
(and thus constant) angular size distance multiplied with the scale
factor, $R\sub b$. This is a common assumption in lensing theory, but
at the level of accuracy that is required to understand the current
tension in cosmological parameters, this simplification may no longer be
sufficient.

World models that are homogeneous on large scales, but have
small-scale deviations near the light paths, have been introduced by
\citet{kantowski69}, \citet{dyer72} and \citet{dyer73}. The basic idea
is that matter is clumped, but clumps near the line of sight are
explicitly described as gravitational lenses, so that the remaining
density around the light rays may be reduced relative to the global
mean, which reduces the focusing and increases the angular size
distances. These inhomogeneities can also be described explicitly as
perturbations at certain redshifts along the line of sight. Even if we
can describe the effect well (e.g.\ if the inhomogeneity parameter in
the Dyer-Roeder distance is known), they break the symmetry of the
homogeneously expanding geometry.

In the following derivations, we typically start with the general
case (described by angular size distances and their derivatives)
before we consider the homogeneous case, in which the time-derivatives
of distance parameters and the prefactor in Eq.~\eqref{eq:timdel} can
be derived as follows, using variants of Eq.~\eqref{eq:dot dt simple}:
\begin{alignat}{2}
  \frac{\diff \ln D\sub{ab}}{\diff t_0} &= \frac{\diff \ln R\sub
    b}{\diff t_0} &&= \frac{H\sub b}{1+z\sub b} \label{eq:dot Dab}
  ,\\ \frac{\diff \ln \Deff}{\diff t_0} &= \frac{\diff \ln R\sub
    d}{\diff t_0} &&= \frac{H\sub d}{1+z\sub d} \label{eq:dot
    Deff},\\ \frac{\diff \ln \Deffcorr}{\diff t_0} &= \frac{\diff \ln
    R_0}{\diff t_0} &&= H_0. \label{eq:dot Deffcorr}
\end{alignat}
In the scenario in which the geometry of the entire Universe,
including the lens, expands homogeneously, the potential and, thus,
the image positions do not change, so that Eq.~\eqref{eq:dot Deffcorr}
describes the only variation with time in Eq.~\eqref{eq:timdel}. This
confirms our general result from Eq.~\eqref{eq:dot dt simple}.

We note that the relation
\begin{align}
  \frac{\diff \ln \Deffcorr}{\diff t_0} - \frac{\diff \ln \Deff}{\diff
    t_0} &= H_0 - \frac{H\sub d}{1+z\sub d}
\end{align}
always holds, even in the inhomogeneous case.

\subsection{Non-expanding lens}
\label{sec:non-exp lens}

We write the Shapiro delay in Eq.~\eqref{eq:psi Shapiro} in terms of a
physical vector, $\vec x=\Dd\,\vec\theta,$ and assume that the function
$\tau\sub d(\vec x)$  is invariant as result of the invariant mass
distribution in physical space. By writing the equation twice, once
for a reference epoch (superscript ref) and once in general but for
the same argument of $\tau\sub d$, we can derive how the lensing
potential $\psi$ of a constant physical mass distribution varies over
time:
\begin{align}
  \psi(\vec\theta) &= S_\psi
  \,\psi\supp{ref}\bigl(S_\theta\,\vec\theta\bigr)
  \label{eq:psi psi ref}
  ,\\ S_\theta &= \frac{\Dd}{\Dd\supp{ref}} & S_\psi &=
  \frac{\Deff\supp{ref}}{\Deff}.
\end{align}
The resulting deflection angle scales such that\ 
\begin{align}
  \vec\alpha(\vec\theta) &= S_\theta S_\psi \,
  \vec\alpha\supp{ref}\bigl(S_\theta\,\vec\theta\bigr) ,\\ &=
  \vec\alpha\supp{ref}\bigl(S_\theta\,\vec\theta\bigr) .
\end{align}
The last form is designated for homogeneous expansion and means that it is
invariant for fixed physical vector, $\vec x$, which is exactly what we expect
for an invariant physical mass distribution.  This relation
automatically holds for an isothermal lens, in which the deflection
angle does not vary in the radial direction. Therefore, such a mass
distribution produces time delays growing with $H_0$, but this
behaviour can only be utilised if we know that the lens is
isothermal. In that situation, we do not need the relative redshifts at
all, but we can determine the Hubble constant directly from the time
delays, for instance, by using the formalism presented by
\citet{wucknitz02}.

For the time derivative of Eq.~\eqref{eq:timdel}, we can neglect the
resulting shift of image positions because they do not affect the
light travel time as result of Fermat's principle. For the
time-derivative of the potential at fixed $\vec\theta$, we have to
consider both scalings in Eq.~\eqref{eq:psi psi ref} and remember that
the gradient is the deflection angle:
\begin{align}
\frac{\partial \psi(\vec\theta)}{\partial t_0} &= \frac{\diff\ln
  S_\theta}{\diff t_0}\, \vec\nabla \psi(\vec\theta) \cdot\vec\theta +
\frac{\diff\ln S_\psi}{\diff t_0}\,\psi(\vec\theta) ,\\ &=
\frac{\diff\ln \Dd}{\diff t_0}\, (\vec\theta-\vec\thetas)
\cdot\vec\theta -\frac{\diff\ln \Deff}{\diff t_0}\,\psi(\vec\theta)
\label{eq:psidot}
  ,\\ &= \frac{H\sub d}{1+z\sub d} \, \Bigl[
    (\vec\theta-\vec\thetas)\cdot\vec\theta - \psi(\vec\theta) \Bigr].
\end{align}
The last equation is designated for the homogeneous Universe.

\subsection{Transverse motion}

It has been known for years that the transverse motion of a gravitational
lens relative to the optical axis has an effect on the observed
redshifts of lensed images. This effect can be interpreted in
different ways; for instance, as a Doppler effect \citep{birkinshaw83},
gravitomagnetic effect of the moving lens
\citep[e.g.][]{pyne93,kopeikin99}, as energy transfer in the
scattering of photons by the lens \citep{wucknitz04}, or directly as
time delays changing with the changing geometry \citep{zitrin18}.

With transverse velocities, $\vec V\sub s$, $\vec V\sub d,$ and $\vec
V_0$, of the source, lens, and observer, we can define the proper motions
\begin{align}
  \dot{\vec\theta}\sub s &= \frac{\vec V\sub s}{(1+z\sub s)\,\Ds}
  , \\ \dot{\vec\theta}\sub d &= \frac{\vec V\sub d}{(1+z\sub
    d)\,\Dd} ,
\end{align}
and write the result of \citet{wucknitz04} for the motion-induced
redshift as
\begin{align}
  c\, z\propmot(\vec\theta) &= \left[ \Deffcorr \left(
    \dot{\vec\theta}\sub d-\dot{\vec\theta}\sub s\right) - \vec
    V_0\right]\cdot\vec\theta .
  \label{eq:z pm}
\end{align}
As a consistency check for the difference between images, we can take
the contribution to the time-derivative of Eq.~\eqref{eq:timdel} due
to the proper motion of the source and confirm that it is consistent
with Eq.~\eqref{eq:z pm} in the case of a fixed lens and
observer. When taking the derivative, we can again neglect the shift
of the images because of Fermat's principle.

For a transverse speed of $300~\kms$, the induced redshift for image
separations of 1~arcsec amounts to $5\times 10^{-9}$, corresponding to
radial Doppler speeds of $1.5~\text{m}\,\text{s}^{-1}$.
\citet{birkinshaw83} had already suggested measuring this effect on
the CMB caused by moving clusters of galaxies, but concluded that it
is difficult. \citet{molnar03} estimated that detecting the effect
with optical spectroscopy is challenging even for massive clusters of
galaxies.  Radio frequencies can be measured with much higher
accuracy, but redshifts can only be determined for sufficiently narrow
spectral features.

With repeating FRBs, we can eventually measure relative redshifts with
a precision on the order of $10^{-14}$ and not by measuring the radio
frequency, but the time lag between bursts, via coherent measurements
of time delays as explained above. This accuracy corresponds to
transverse speeds of only about one metre per second or proper motions
on the order of $10^{-13}~$arcsec per year. Later in this paper, we  demonstrate how well
the proper motion can be disentangled from the various other effects
on the redshifts.

We note that in the equation for the scaling of the lensing potential
of a non-expanding lens in Eq.~\eqref{eq:psi psi ref}, we implicitly
assumed that the centre of the scaling law is the coordinate
origin. An offset in this centre is equivalent to an additional proper
motion that corresponds roughly to the offset over a Hubble time.

\subsection{Combining delays and redshifts}

We now try to combine the equations for the time delays,
Eq.~\eqref{eq:timdel}, with the ones for its derivatives (or
redshifts), including the contribution from the non-expanding lens,
Eq.~\eqref{eq:psidot}, and the transverse motion of the source from
Eq.~\eqref{eq:z pm}. For simplicity, we assume that lens and observer
are at rest, which is still fully general, because only relative
motion matters. For each image, we have one equation for the time delay
and one for the redshift:
\begin{align}
  \Delta t_0 &= \frac{\Deff'}{c} \left[ \frac{\vec\theta^2}{2} -
    \vec\theta\cdot\vec\thetas - \psi(\vec\theta) \right] +
  \text{const}
\label{eq:timdels}
  ,\\ z(\vec\theta) &= \frac{\Deff'}{c} \, \Biggl\{
  \frac{\diff \ln \Deffcorr}{\diff t_0} \left[\frac{\vec\theta^2}{2} -
    \vec\theta\cdot\vec\thetas - \psi(\vec\theta)\right]
  +\frac{\diff\ln \Deff}{\diff t_0}\,\psi(\vec\theta) \notag \\ &
  \qquad - \frac{\diff\ln \Dd}{\diff t_0}\, (\vec\theta-\vec\thetas)
  \cdot\vec\theta - \dot{\vec\theta}\sub s \cdot \vec\theta \Biggr\} +
  \text{const}
\label{eq:redshifts}
  .
\end{align}
If we consider only the differences between images, we can eliminate the
additive constants. A lens system with $n$ images provides $n-1$
equations each of type, \eqref{eq:timdels} and
\eqref{eq:redshifts}. The unknowns are $n-1$ potential differences,
two components of the source position, two of the proper motion, and
some number of cosmological parameters. Even in a quadruply lensed
system there are not enough constraints to determine all unknowns, not
even if the cosmology is known.
Fortunately, as we  see in the text below, nature is so kind that it allows us to
eliminate all unwanted parameters and determine the more interesting
ones.

Because the potential is the major uncertainty in classical
applications of gravitational lensing for cosmology, we substitute the
time delay for both occurrences of the potential in
Eq.~\eqref{eq:redshifts}, which is trivial for the term in square
brackets. We are then left with equations for the redshifts with only
cosmological and geometrical terms:
\begin{align}
  z(\vec\theta) &= Z_t \,\Delta t_0 + Z_\theta\, \frac{\theta^2}{2}
  -\frac{\Deffcorr}{c} \,\dot{\vec\theta}\sub s' \cdot \vec\theta +
  \text{const}
   \label{eq:z with delta t}
  ,\\ Z_t &= H_0 - \frac{H\sub d}{1+z\sub d}
    \label{eq:def Z t}
  ,\\ Z_\theta &= \frac{\Deffcorr}{c} \left( \frac{\diff
    \ln\Deff}{\diff t_0} - 2\frac{\diff \ln\Dd}{\diff t_0} \right)
  \label{eq:def Z theta}
  ,\\ \dot{\vec\theta}\sub s' &= \dot{\vec\theta}\sub s
  +\left(\frac{\diff \ln\Deff}{\diff t_0} - \frac{\diff \ln\Dd}{\diff
    t_0} \right) \vec\thetas.
  \label{eq:def pm prime}
\end{align}
We eliminated the unknown lensing potential and find that the position
and proper motion of the source only appear in the combination
$\dot{\vec\theta}\sub s'$. Because we have to determine or eliminate
the proper motion in any case, the unknown source position does not add
additional free parameters.  When studying the proper motion itself, the
deviation in $\dot{\vec\theta}\sub s$ is still small, for galaxy-scale
lenses it is typically $\ll 1~\kms$.

Cosmological parameters can be determined or constrained via the
proxies $Z_t$ and $Z_\theta$.

In the homogeneous case, we can now apply Equations \eqref{eq:dot Dab}
and \eqref{eq:dot Deff} and find
\begin{align}
  Z_\theta &= -\frac{H\sub d\,\Deff}{c}
\label{eq:z with delta t spec}
  , \\ \dot{\vec\theta}\sub s' &= \dot{\vec\theta}\sub s
  .
\end{align}
For an isothermal lens, the potential can be eliminated even without
redshifts and Eq.~\eqref{eq:timdels} is reduced to
\begin{align}
    \Delta t_0 &=- \frac{\Deff'}{c}\frac{\vec\theta^2}{2} +
    \text{const} ,
\end{align}
as described by \citet{wucknitz02}. With this, the two terms with
$H\sub d$ in Eqs.~\eqref{eq:def Z t} and \eqref{eq:z with delta t
  spec} cancel out and we are left with a pure dependence on $H_0$ and
no other cosmological parameters. Unfortunately, this only helps if we
know about the isothermality.

After eliminating the lensing potential, we are left with effectively
$n-1$ equations from the redshifts less the additive constant.
Unknowns are the proper motion and cosmological parameters.  For a
double-imaged system, we cannot constrain the cosmological parameters, but
we can still estimate one component of the proper motion with slightly
reduced accuracy.  The terms with $Z_t$ and $Z_\theta$ in
Eq.~\eqref{eq:z with delta t} are on the same order of magnitude,
about $2\times10^{-12}$, as shown in Table~\ref{tab:numbers}. Because the
proper-motion-induced redshift is much larger, typically
$5\times10^{-9}$, we can neglect the cosmological terms for a
measurement of the proper motion component along the image separation
with an accuracy on the order of $0.1~\kms$, but we have to know the
distance parameters to convert it to physical units.

For a quad system, we can determine the full proper motion vector in
the same way and, in addition, obtain constraints on the cosmological
parameters. We can treat the additive constant in Eq.~\eqref{eq:z with
  delta t} as a free parameter, invert the set of equations, and
express our two `cosmological parameters' $Z_t$ and $Z_\theta$ as well as the
proper motion as known linear functions $f_1$, $f_2$, $\vec f_3$ of
the constant:
\begin{alignat}{2}
  f_1(\text{const}) &= Z_t &&= H_0 \left[1 - \frac{H\sub d}{(1+z\sub
      d)\,H\sub 0}\right]
 \label{eq:f1}
  ,\\ f_2(\text{const}) &= -Z_\theta &&= \frac{H\sub d}{H_0} \deff
 \label{eq:f2}
  ,\\ \vec f_3(\text{const}) &= \frac{\Deffcorr}{c}\dot{\vec\theta}\sub
  s' & \quad&= \frac{\deffcorr}{H_0} \dot{\vec\theta}\sub s
. \label{eq:f3}
\end{alignat}
Here, the parameters are also written for the homogeneous case in an
alternative way that separates the effect of $H_0$ and the other
cosmological parameters. For this, we introduced the reduced distance
parameter, $\deff=H_0 \Deff/c$, which is independent of $H_0$. If the
other parameters are known, $H_0$ and $\dot{\vec\theta}\sub s$ can be
determined uniquely. Otherwise, the information from several lensed
FRBs can be combined to determine all parameters (see
Sec.~\ref{sec:several lenses}).

These equations show explicitly that the two proxy cosmological
parameters, $Z_t$ and $Z_\theta$, and the proper motion can be expressed
as linear functions of a free parameter. This means that one lens
system provides only one effective constraint on cosmology, for
instance, on the Hubble constant if all other parameters are
known. More is possible by combining several lenses, as we show
in later sections.
  
Even in the inhomogeneous case, we have good constraints on cosmology,
but slightly more obscure. Eq.~\eqref{eq:f1} remains unchanged, but
the right hand sides of Eqs.~\eqref{eq:f2} and \eqref{eq:f3} have to
be replaced by their more general versions by using Eqs.~\eqref{eq:def Z
  theta} and \eqref{eq:def pm prime}.

\subsection{Radial motion}

The radial motion of source, lens, and observer impacts the result for
two reasons. Firstly, there is the direct effect on the time derivative
of distances. In Euclidean geometry, we would have $\dot D\sub{ab} =
v\sub b - v\sub a$. This simple relation does not hold for
cosmological distances, but the order of magnitude would be
similar. The characteristic $\dot D/D\approx v/D$ is on the order of
$10^{-21}~\text{s}^{-1}$ for 1~Gpc and $300~\kms$. Compared to the
Hubble constant, this is below one percent, so that the direct effect
of radial motion is not of concern here, although it should
be investigated further.

The second effect of relevance is relativistic aberration, which
shifts the apparent positions of sources towards the apex of
motion. When approaching a source, it appears compressed, which can be
interpreted as an increased angular size distance. To first order we
find the corrected distance,
\begin{align}
  \tilde D\sub{ab} &= \left(1+\frac{v\sub a}{c}\right) D\sub{ab}
  ,
\end{align}
which depends on the motion of the observing side only.\footnote{Of
  course, physical effects should only depend on relative motion. A
  situation with a moving source can be Lorentz-transformed to that of
  a moving observer. The relativistic length contraction does not
  matter here because it is of second order. More important is the
  effect of redefined simultaneity. In one situation, we define the
  distance at the moment of emission and in the other at the moment of
  observation. The relative motion between these two moments accounts
  for the difference in the two descriptions.}  For non-relativistic
velocities, this is an insignificant scaling. The effect on the derived
proper motion is larger than its formal precision, but still
remains rather small. Radial motion also modifies the observed redshifts, which are
needed to calculate distances given a cosmological model. This is a
small effect that we can neglect.

We conclude that uniform radial motion generally would not require
significant corrections. A realistic acceleration, however, can have a
relevant effect on time-derivatives of the distances and, therefore, on
the observed redshifts. For the additional contribution to the
time derivative of the time delays in Eq.~\eqref{eq:timdel}, we need
the effect on the effective distance, to the first order in the velocities
\begin{align}
\tildeDeff &= \left(1+\frac{2v_0-v\sub d}{c}\right) D\sub{eff}
,\\ \left.\frac{\diff \ln \tildeDeff}{\diff t_0}
\right|\sub{accel} &= \frac{2 a_0}{c} - \frac{a\sub d}{(1+z\sub d)\,c}
.
\end{align}
Here, the accelerations, $a_0$ and $a\sub d$, are measured in their
respective comoving frames and the latter had to be scaled for
redshift to obtain the derivative of $v\sub d$ with respect to
the observed time.
Formally, we also have to consider that velocities relative to the
Hubble flow decrease due to the expansion,\footnote{This is not a
  physical effect in the sense of a deceleration but because
  the object moves into a region with a slightly different Hubble
  flow, the relative speed is reduced.} but this effect is
smaller than the Hubble expansion itself by a factor of $v/c$ and,
therefore, negligible.

For the observer, we assume that the diurnal motion and orbit around
the Solar system barycentre have been corrected. Another relevant
effect is the Galactocentric acceleration. As we argue in
Section~\ref{sec:earth motion}, these effects can be corrected for with
sufficient accuracy.

The radial acceleration of the lensing galaxy is potentially more
critical because it cannot be measured independent of the lensing
redshifts and, thus, it acts as additional source of (probably highly
non-Gaussian) statistical errors. For the Milky Way, the acceleration
within the local group is dominated by the gravitation of the Large
Magellanic Cloud. With a total mass of $1.4\times 10^{11}~\msun$
\citep{erkal19} and a distance of 50~kpc, the gravitational
acceleration divided by $c$ is $2.6\times 10^{-20}~\text{s}^{-1}$,
which corresponds to about one percent of the Hubble constant. If this
is typical for lensing galaxies, the effect on the achievable accuracy
is limited, but a good sample of lens systems should be used to
identify possible outliers.

Finally, we see that radial acceleration of the source has no effect on
the derived cosmological parameters. This is essential, because values
similar to the Galactocentric acceleration or even much larger have to
be expected, which would otherwise invalidate the method. Very high
velocities of the source can still be relevant due to the direct
effect on time derivatives of distance parameters. High velocities are
expected in tight orbits around massive objects, which may hopefully
be identified by their time dependence. In the unlikely case that FRBs
are emitted from relativistically moving systems, for example, in AGN jets,
their motion would swamp the effects that we are interested in. This
curse for cosmology would be a blessing for FRB physics and, thus, time
delays and relative lensing redshifts could be used to study the FRB
motion with the highest precision.

\section{Cosmology with ensembles of lensed FRBs}
\label{sec:several lenses}

\begin{figure}
\includegraphics[width=\hsize]{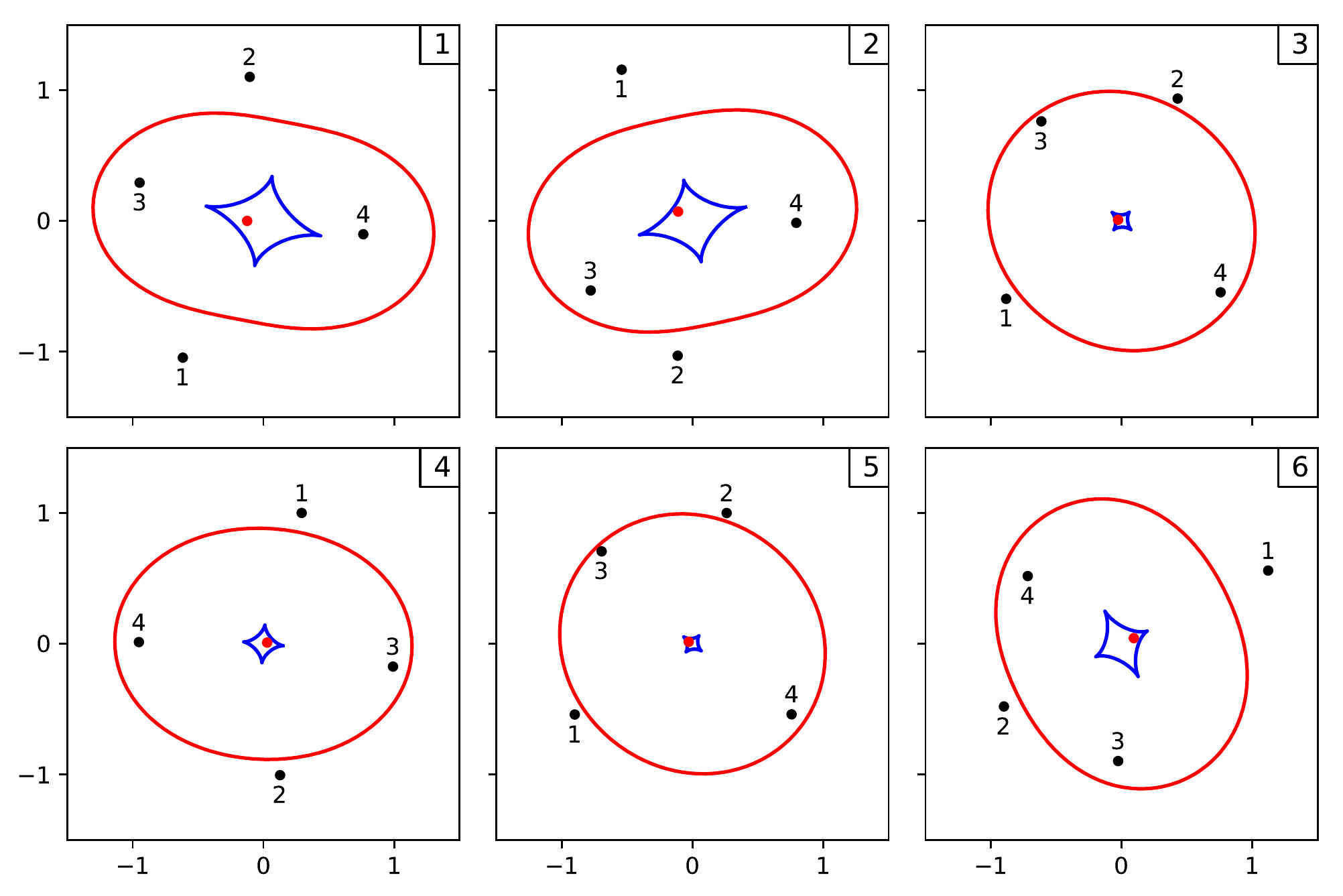}
  \caption{Geometry of the six lenses in our mock sample. The outer
    red curves are the critical curves and the inner blue diamond-shaped
    curves are the caustics. Images are shown as black dots labelled
    by increasing time delay and the true source positions are the red
    dots near the centres. The scale is given in arcseconds.}
  \label{fig:lenses}
\end{figure}

\begin{table}
  \caption{Uncertainties (standard deviations) of $H_0$ and $\Omegam$
    determined individually (assuming the other and $w$ known) from
    the six lenses. The posterior distributions are very close to
    Gaussian with mean very near the `true' values of $H_0=70~\kmsmpc$
    and $\Omegam=0.3$.}
  \label{tab:MCMC H0 Om}
  \centering
  \begin{tabular}{ccc}
    \hline\hline lens & $\sigma_{H_0}~[\kmsmpc]$ & $\sigma_{\Omegam}$
    \\ \hline 1 & \ph01.38 & 0.0125 \\ 2 & \ph01.49 & 0.0072 \\ 3 &
    \ph07.87 & 0.0279 \\ 4 & \ph03.13 & 0.0198 \\ 5 & 15.67 & 0.0593
    \\ 6 & \ph02.44 & 0.0071 \\ \hline
  \end{tabular}
\end{table}

We explain in the analysis above how one gravitationally lensed
repeating FRB provides one constraint on the combination of proxy
cosmological parameters $Z_t$ and $Z_\theta$.
If all cosmological parameters with the exception of the Hubble
constant are known, $Z_\theta$ can be calculated and $Z_t$ is simply
proportional to $H_0$ with a known scale factor, so that we have a
direct measurement of the Hubble constant. Using our reference numbers,
we expect an accuracy of 0.5\,\%.  In reality, we have to take into
account the accuracy of the image positions and how accurately the
system of equations can be solved for $H_0$, which will depend on the
image configuration of the lens.

For this first investigation of the potential as cosmological probes,
we are not attempting to define a truly representative sample of lensed
FRBs -- which is currently impossible given that we do not even know the source
redshift distribution. Instead, we use a rather arbitrary sample of
lenses with isothermal elliptical potentials plus external shear, all
with the same Einstein radius $E=1~$arcsec.\footnote{We emphasise that
  the knowledge of this restriction is not used when deriving
  cosmological parameters from the simulated data. This is different
  from fitting isothermal models.} Adding an external convergence or
using a range of power-law indices would generally make the results
better by providing more diversity, which reduces degeneracies. Having
external shear and ellipticity, or at least some lenses with shear and
some with ellipticity, also turns out to be important, which has to be
investigated further.

The potential for our simple family of models is given by
\begin{align}
  \psi (x,y) &= E
  \sqrt{\frac{x^2}{(1+\epsilon)^2}+\frac{y^2}{(1-\epsilon)^2}}
  -\frac{\gamma_1}{2}(x^2-y^2) - \gamma_2\,x\,y ,
\end{align}
in which $(\gamma_x,\gamma_y)$ represents the external shear and
$\epsilon$ the ellipticity of the potential. For small ellipticities,
this is a good approximation of an elliptical mass distribution.
The parameters were chosen from Gaussian distributions and the source
positions from a uniform distribution within the central caustic to
produce quadruple images. For the simulations, we assume a spatially
flat homogeneous cosmological model with $H_0=70\,\kmsmpc$, matter
density $\Omegam=0.7$, dark-energy density $\Omegaw=0.3$ and
equation-of-state parameter $w=-1$, equivalent to a cosmological
constant.  This cosmology is used to calculate, from the lens models in
angular coordinates, the time delays and their time
derivatives.\footnote{{\tt AstroPy} \citep{astropy13,astropy18} was
  used for the cosmological calculations and results were checked
  against our own integration code that is more general and can also
  compute Dyer-Roeder distances for inhomogeneous models.} The image
configurations are illustrated in Fig.~\ref{fig:lenses} and parameters
are listed in Table~\ref{tab:lenses}. With the exact values as mock
measurements we use Markov Chain Monte Carlo (MCMC) simulations to
explore the parameter space that is consistent with these
measurements. Practically we use the {\tt MultiNest} software
\citep{feroz09} and its {\tt Python} interface
\citep{buchner14,buchner16}.

The only explicit free parameters for the MCMC simulation are the
cosmological parameters. The image positions are fitted implicitly
within the loop by analytically minimising the deviations of the image
positions and the redshifts. This $\chi^2$ is then used for the
likelihood of the outer MCMC loop. Tests show that this accelerated
approach introduces only negligible errors. The time delays themselves
can be used directly because of their extreme precision. Their
uncertainty matters only in so far as the time delays are used to
determine the relative redshifts.

\begin{figure*}
  \centering \includegraphics[width=0.8\hsize]{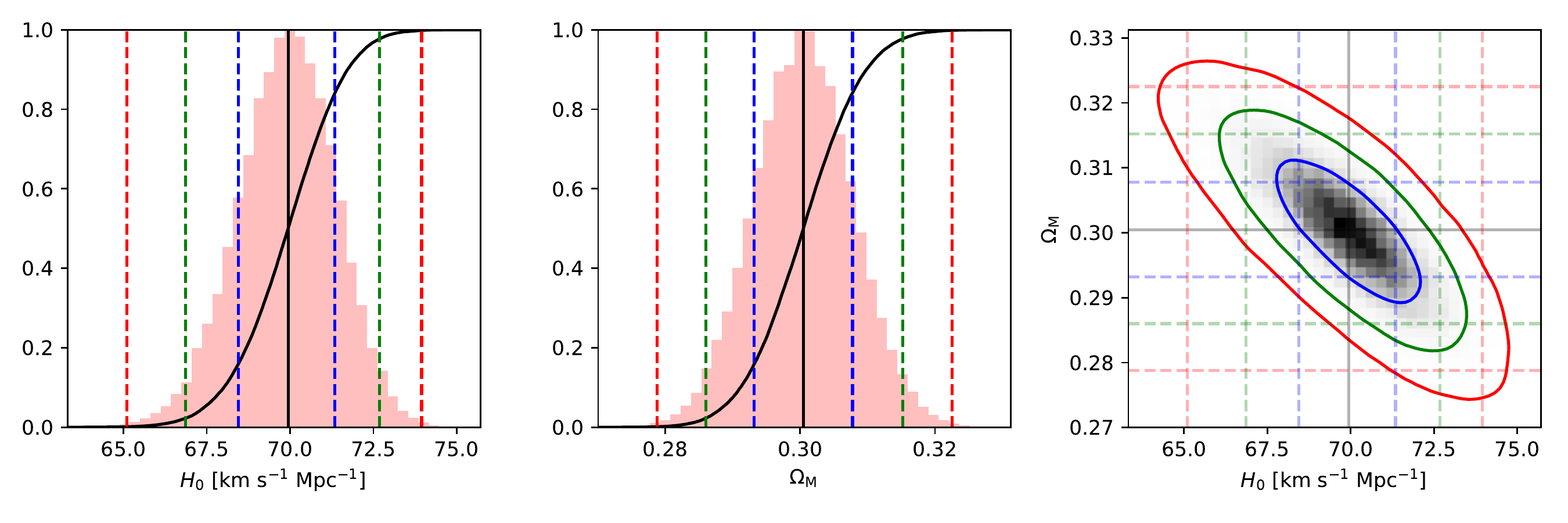}
  \caption{Posterior distribution of $H_0$ and $\Omegam$ using all
    lenses simultaneously. Two left panels show the marginalised
    distributions. Vertical dashed lines are the 1,2,3 $\sigma$ limits
    (68.3\,\%, 95.4\,\%,\, 99.7\,\%, respectively, in blue, green, red)
    and the median (black solid). The total plot range is adapted to
    the 3 $\sigma$ range.  The right panel shows the two-dimensional
    distribution. The marginalised limits are included as shaded
    horizontal and vertical lines. The contours show the 2-dimensional
    1,2,3 $\sigma$ limits, respectively, in blue, green, red. The
    total range is the same as in the left panels. }
  \label{fig:MCMC cosmo 2}
\end{figure*}

\begin{figure*}
  \centering
  \includegraphics[width=0.83\hsize]{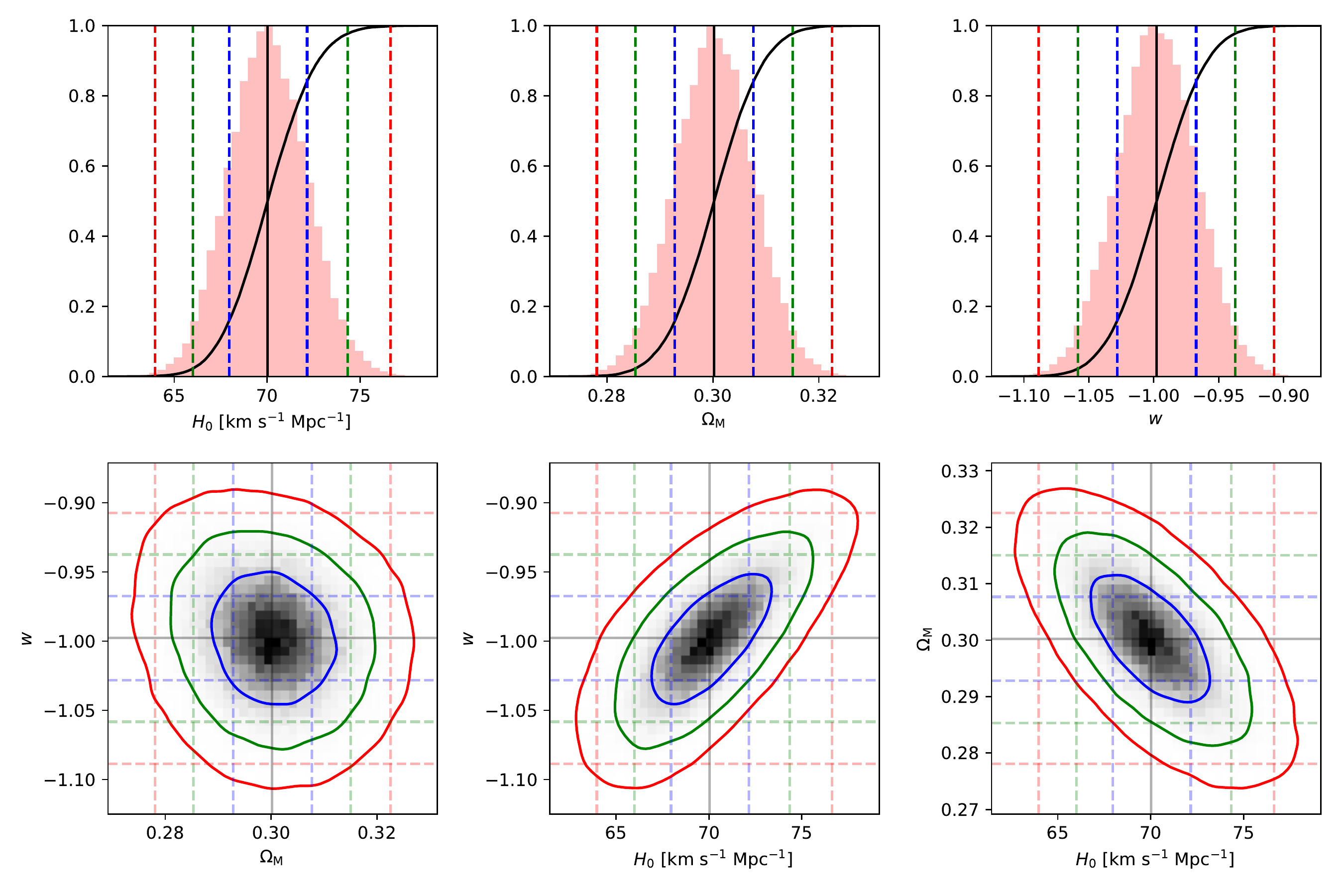}
  \caption{Posterior distribution of $H_0$, $\Omegam$ and $w$ using
    all lenses simultaneously. The upper row shows marginalised
    distributions for individual parameters, the lower row for all
    combinations of two. See Fig.~\ref{fig:MCMC cosmo 2} for a
    detailed description.}
  \label{fig:MCMC cosmo 3}
\end{figure*}

As our measurement uncertainties, we assume $10^{-14}$ for the relative
redshifts and 0.5~mas for the image positions (0.35~mas for each
component). It so happens that these errors influence the resulting
uncertainty to a very similar extent, so that both have to be reduced
for a significant improvement.

Firstly, we investigate how well individual parameters can be determined
from individual lenses, assuming that all others are
known. Table~\ref{tab:MCMC H0 Om} shows the resulting uncertainties
for the Hubble constant and the matter density, assuming the other is
known and a spatially flat Universe with $w=-1$.  We find that the
uncertainties vary strongly from lens to lens, which means that more
realistic samples should be investigated in the future. Generally, we
find that more asymmetric lenses (visible as larger caustics) are
better, but the details have to be worked out.

Because a measurement of the Hubble constant alone is not equipped to
solve the current parameter tension, we also simulate fits of sets of
parameters to the full ensemble of lenses. Figure~\ref{fig:MCMC cosmo 2}
and Table~\ref{tab:MCMC cosmo 2} show results for the Hubble constant
and matter density for a flat Universe with $w=-1$.
Figure~\ref{fig:MCMC cosmo 3} and Table~\ref{tab:MCMC cosmo 3} add $w$ as
free parameter. Some of the correlations between these parameters are
significant, but none are so extreme to preclude the determination of
all. This may change if we relax the condition of a spatially flat
Universe.

The potential for the achievable accuracy with this approach is
competitive with that achievable by combinations of other methods,
particularly with regard to the matter density and the equation of
state of dark energy. It remains to be seen how realistic our
assumptions are.

\section{Caveats}
\label{sec:caveats}

There are a number of effects that can perturb the measured time
delays and (more importantly) their time-derivatives. Most relevant is
the proper motion, which can be determined and implicitly corrected
for quad systems as shown above. The accuracy of this correction
depends on the astrometric accuracy. The radial motion is also discussed
above, where we note that the motion of the Earth relative to the
background metric must be taken into account.

\begin{table}
  \caption{Uncertainties (standard deviations) of $H_0$ and $\Omegam$
    determined from all six lenses combined for a spatially flat
    Universe with $w=-1$. The right two columns show the correlation
    coefficients between the parameters.}
  \label{tab:MCMC cosmo 2}
  \centering
  \begin{tabular}{lc@{\qquad}cc}
    \hline\hline & & \multicolumn{2}{c}{correlation with} \\ parameter
    & $\sigma$ & $H_0$ & $\Omegam$ \\ \hline $H_0~[\kmsmpc]$ & 1.46 &
    1 & $-0.789$ \\ $\Omegam$ & 0.0073 & $-0.789$ & 1 \\ \hline
  \end{tabular}
\end{table}

\begin{table}
  \caption{Uncertainties (standard deviations) of $H_0$, $\Omegam,$ and
    $w$ determined from all six lenses combined for a spatially flat
    Universe. The right three columns show the correlation
    coefficients between the parameters.}
  \label{tab:MCMC cosmo 3}
  \centering
    \begin{tabular}{lc@{\qquad}ccc}
    \hline\hline & & \multicolumn{3}{c}{correlation with} \\ parameter
    & $\sigma$ & $H_0$ & $\Omegam$ & $w$ \\ \hline $H_0~[\kmsmpc]$ &
    2.09 & 1 & $-0.650$ & $+0.707$ \\ $\Omegam$ & 0.0074 & $-0.650$& 1
    & $-0.138$ \\ $w$ & 0.030 & $+0.707$ & $-0.138$ & 1 \\ \hline
  \end{tabular}
\end{table}

\subsection{Motion of the observer}
\label{sec:earth motion}

As explained above, we have to correct the arrival times of bursts for
a number of effects. Firstly, there is the diurnal motion of the observer
(which is extremely well-known) and the Earth's orbit around the Solar
system barycentre. Relative to the known planets and the Sun, this
motion is certainly known to better than 300~m, which would correspond
to the claimed timing accuracy of one microsecond. Unfortunately we
need to know the motion relative to a local inertial frame. Should
there be additional large unknown masses in the Solar system, for
instance, an unknown distant planet, its direct influence on the motion
of the inner planets may be small, but it does introduce a relative
motion between the true barycentre and the one derived from the known
planets.

With timing observations of a binary pulsar, relative radial
acceleration between the pulsar and us can be measured very
accurately. \citet{verbiest08} did this for PSR J0437--4715 with an
accuracy corresponding to $a/c$ of $1.2\times 10^{-19}~\text{s}^{-1}$,
which is about 5\% of the Hubble constant. Within the measurement
accuracy, the result is consistent with the Galactocentric acceleration
and kinematic effects, with a remaining uncertainty of any unmodelled
contributions of $6\times 10^{-19}~\text{s}^{-1}$. This constraint was
improved by \citet{deller08} with a new measurement of the parallax,
with a limit of $1.6\times 10^{-19}~\text{s}^{-1}$ at the 2-$\sigma$
level. The acceleration measurement itself was also improved by
\citet{reardon16} by an order of magnitude and is now at an
uncertainty of $1.2\times 10^{-20}~\text{s}^{-1}$. It is still
consistent with the Galactic and kinematic contributions.

For our acceleration relative to an inertial frame, we have to
consider the total Galactocentric acceleration of the observer. This
amounts to $a/c=8\times10^{-19}~\text{s}^{-1}$ or about one third of
the Hubble constant. It is thus a very relevant effect that must be
corrected for. \citet{macmillan19} show an overview of results from
Galactic kinematics compared with VLBI measurements via the resulting
apparent proper motion of extragalactic sources. These measurements
include the acceleration of the Milky Way towards the average Hubble
flow. Their accuracy (on the order of $6\times 10^{-20}~\text{s}^{-1}$)
is currently sufficient to correct lensing redshift measurements for
most of the effect, but further improvements are required.

With a sufficiently large sample of lensed FRBs, our local
acceleration can be measured (and corrected) from the observed lensing
redshifts because its effect is position dependent. This can
potentially even be used to improve the Solar system ephemeris, to find
unknown masses in the outer Solar system, and to study the
acceleration of the Milky Way relative to the background.  This
approach carries the disadvantage that it can absorb potential large-scale
anisotropies, which may then go unnoticed.

\subsection{Evolution of lens structure}

Our entire argument is based on the assumption that the mass
distribution of the lens stays constant with time. According to
Table~\ref{tab:numbers} we are measuring relative changes of time
delays on the order $2\times10^{-10}$, corresponding to substantial
changes over time scales of 10~Gyr, which means that even small
violations of the assumption will matter. Typical rotational time
scales are much smaller, but they are not relevant as long as the
rotating mass distribution is symmetric. Major encounters or mergers
will certainly introduce large perturbations, but we hope that such
systems can be excluded based on their morphology.
\citet{walker96} discusses how gravitational lensing by moving stars
in the Milky Way might influence pulsar timing, discussing effects that
are almost always too small to be relevant.

Other changes are still of concern and have to be investigated in the
future using structure formation simulations. These not only
have the potential to provide the mass distribution at a given moment, but also the changes
over time. Even if they do not reproduce realistic mass distributions
of galaxies exactly, they will allow us to estimate typical effects on
observed lensing-induced differential redshifts.

\subsection{Masses near the line of sight}

Our approach of combining time delays and their time derivatives
(redshifts) allows us to eliminate the unknown mass distribution of
the lens and break the mass-sheet degeneracy, as long as all unknown
masses are at the same known redshift. Additional over- or
underdensities along the line of sight will not automatically be
corrected for. Because even though their total static effect is usually small
\citep[e.g.][]{millon19}, we expect that the remaining error in our
method will not introduce large systematics, but this has to be
investigated in detail.

More worrying are the potential effects on the time derivatives due to
moving or evolving masses near the line of sight. Their influence has
to be studied, again, likely based on simulations of the large-scale
structure formation.

Masses outside of the main lens will have only a small effect. The
zeroth-order (in position) redshift caused by the evolving lensing
potential does not effect the relative redshifts. The first order is expected to
be much lower and  absorbed in the proper motion, for which it
introduces only a small error. Only higher orders remain, which are
sufficiently low for masses that are not too close to the line of
sight. It is thus expected that most lens systems should not be
problematic, but this has to be worked out quantitatively.

\subsection{Astrometry of lensed images}

The ability to measure and correct for the relative proper motions
relies on accurate measurements of the relative positions of the
lensed images, with uncertainties on sub-mas levels. For bright
persistent sources, for instance, lensed AGN, this is not
problematic. With VLBI at L-band near 1.4~GHz we can achieve
resolutions of a few mas, even better at higher frequencies. The
achievable positional uncertainty scales with the resolution divided
by the S/N, such that relative positions better than 0.1~mas, at least
for pointlike sources, are routinely achieved.

For FRBs, the situation is more difficult. Firstly, they are short, so
that the S/N will generally be lower than for AGN sources. With
careful gating (S/N-based weighted averaging), it is still realistic to
achieve an S/N of 100, which would be
sufficient.\footnote{\citet{marcote20} find bursts with SNR of about
  50 for the 100-m Effelsberg telescope, even without optimal gating
  of the small-scale structure.} Another difficulty is the very poor
image fidelity due to the sparse UV coverage for individual
bursts or images. Even with proper model-fitting to the observed
visibilities, this causes highly non-Gaussian errors, which can only
be beaten down by combining several bursts.

Finally, we can only see one image at a time, so the positions
cannot be measured directly relative to each other, but have to be
relative to a nearby calibrator source using phase-referencing. This
always leaves residual errors due to the atmosphere and ionosphere. In
principle, these can be reduced by observing a number of reference
sources around the target, but if they are not within the same primary
beam of the telescopes, we have to alternate scans of target and
references, which means we may miss individual bursts. Alternatively, we
can point at the target and detect bursts in realtime to move to
calibrators directly after a burst. This, however, is not a standard
VLBI observing mode.

For the first repeating FRB, \citet{marcote17} achieved an accuracy of
2--4~mas per axis by combining four bursts. The second
VLBI localisation of another FRB by \citet{marcote20} had a similar
accuracy even for individual bursts. There is certainly room for
significant improvement, but the astrometry will nevertheless be
challenging.

In addition to VLBI, we can also use the timing of the lensed burst images
 year-round to measure or improve the astrometry, which is similar to the
technique for determining positions from pulsar timing. If we have a
frequently repeating FRB with good coverage along the Earth's orbit
around the Sun, the effects of the orbit can be disentangled from the
long-term differential redshifts. The orbit itself is then a lever arm
of 1~AU. With a timing accuracy of one microsecond,\footnote{We repeat
  that this accuracy is not available for lags between bursts, but
  only for lensing delays between images. If the time delays are a
  good fraction of a year, we still have very accurate measurements
  between different Earth positions, which is sufficient for a good
  localisation.} this corresponds to a positional accuracy of about
0.4~mas per measurement, unaffected by the atmosphere or
ionosphere. The timing measurement requires much less effort than a
full global VLBI experiment, so that it can be repeated for very many,
maybe thousands of bursts, with the corresponding increase in
precision.

\subsection{Microlensing}

We have to distinguish between two types of microlensing. The first is
due to individual masses along the line of sight, as analysed by
\citet{chang79}. Even if a large fraction of dark matter exists in the
form of compact masses, the chance of lensing is small for any given
line of sight and our proposed method will generally not be
affected. This type of microlensing in FRBs has already been proposed
to study the abundance of compact objects by \citet{munoz16}, who
argue that masses in the range 10--100~$\msun$ can be detected or
constrained using incoherent methods. Using spectral features in the
emission caused by (coherent) interference between the lensed images,
\citet{eichler17} describe how the range probed can be extended to
much lower masses. The typical order of magnitude of the time delay
between lensed images of a point-mass, assuming a modest amplification
ratio, is given by the light-crossing time of the Schwarzschild
radius, which is 10~$\mu$s for one Solar mass and scales proportionally
to the mass.

What is potentially more problematic for our application is microlensing
caused by the dense ensemble of masses in the lensing galaxy. Because
the density must be high for macrolensing to occur, macro-images will
always be affected by some degree of microlensing. Microlensing by the
star field can produce a high number of subimages but, overall, only
a few of them contribute significantly to the total flux
\citep{saha11}. The probability distribution of this number also
depends on the type of image (minimum, maximum, or saddle-point of the
arrival-time function). For known (non-FRB) lens systems, the
micro-images cannot be resolved, not even with VLBI observations of
radio lenses. The only observed signature is the total amplification,
and its variation seen in light curves. Astrometric effects are slowly
getting in reach now for the GAIA satellite.

Time delays in this type of microlensing have not been investigated
thus far. Our preliminary simulations indicate that typical time
delays between the brighter micro-images are on the same order of
magnitude as the corresponding single-lens time delays, or larger by a
factor of only a few. For a Solar mass, thus we have to expect delays
in the range of a few times 10~$\mu$s. This is well above the assumed
accuracy of one microsecond (Tab.~\ref{tab:numbers}) and thus
potentially detrimental for our proposed cosmological application of
(macro-) time delays. Luckily the situation is not hopeless. What is
relevant for our purposes is not so much the accuracy of the absolute
time delays, but the accuracy of their rate of change. For a rough
estimate of the typical time scale, we can divide the Einstein radius
of typical masses (a few times $10^{11}~$km, angular scale a few
$\mu$arcsec) by the typical transverse velocity, which results in many
decades. Because we were assuming to measure changes over
approximately three years, the effect is reduced by at least
an order of magnitude, which brings it into an acceptable range.

We can measure the wave field, which enables us to distinguish the
microlensed images via their time delays. Generally, we expect there to only be a
few relevant subimages, so that we can monitor their relative
separations over the years and use this data to estimate the possible error
on the derived macro time delays. The micro time delays can be
measured with nanosecond precision, which provides invaluable and
entirely new information about the mass distribution and motion of
compact objects in the lensing galaxies. In contrast to optical
microlensing studies, in which most constraints are degenerate with
the source size, we can directly measure micro time delays (which are
a measure for the masses) and their changes over time (which are a
measure of their kinematics).
Both the potential for dark-matter studies and the possible
problematic influence on the cosmological application have to be
studied in detail in the future.

\subsection{Interstellar scattering}

The interstellar plasma has a refractive index for radio waves that is
proportional to the electron density and to $\lambda^2$. Density
fluctuations cause deflections that are similar in nature to
microlensing and can produce a high number of subimages that interfere
with each, which we observe as interstellar scintillation.  The effect
is independent for the lensed macro-images, so that it spreads out the
correlation between the measured wave fields and reduces the accuracy
of a macro-delay determination. Sophisticated methods have been
developed to study the effect in scintillating pulsars, but these rely
on a good time coverage in order to model the scattering field. For
FRBs we do not have this luxury, but have to try and reduce the effect
as much as possible by observing at high frequencies. Scattering sizes
roughly scale with $\lambda^{2.2}$, corresponding time delays with
$\lambda^{4.4}$.

The range of scattering delays depends critically on the line of
sight. For scattering within the Milky Way, low Galactic latitudes
generally show much stronger effects. Using the NE2001 model for the
distribution of free electrons in the Milky Way \citep{cordes02}, we
find that for 1.4~GHz the scattering delay is always below 1$\mu$s for
Galactic latitudes above $\pm 20\degr$, and mostly even above $\pm
10\degr$, so that the redshift accuracy will stay within our assumed
limits for most of the sky.
In known FRBs, it has been determined that the intergalactic medium generally adds
much less scattering than the Milky Way. Scattering in the host is
extremely variable. By finding FRBs at low frequencies, for instance
with CHIME, those with low scattering in the host are automatically
selected.

An additional important factor in lensed FRBs may be the scattering in
the lensing galaxy itself, or in the intergalactic or intracluster medium
around it. If the lens scatters as strongly as the disk of the Milky
Way, the range of scattering delays would indeed make a delay
measurement with the required accuracy impossible. To our advantage,
the most efficient gravitational lenses are massive elliptical
galaxies. As result of their old population of stars and the
corresponding absence of H{\sc ii} regions and supernova remnants,
their already thin interstellar medium is also calm without strong
fluctuations in the plasma density. They are thus expected to have
levels of scattering that are orders of magnitude lower, if
diffractive scattering can happen at all. Admittedly, the physical
processes involved in scattering are not even fully understood in the
Milky Way and the extrapolation to elliptical galaxies might turn out
to be too optimistic.
  
Because of the strong frequency dependence, we can always reduce the
effect by going to higher frequencies. This will generally be
necessary if we want to study the delays of micro-images in detail.
If, in the worst case, the scattering is too strong to measure time
delays with accuracies of a microsecond, we will notice immediately
because no sharp peak will be found in the coherent correlation. The
accuracy can be estimated from the width of the peak and there is no
danger that an unexpected large error would go unnoticed.

\subsection{Ionospheric and atmospheric delays}

These effects are relevant for VLBI-astrometry of the lensed images, but can
be neglected for the redshifts derived from measured time delay
changes. The troposphere has an effect of less than 10 nanoseconds at
zenith and most of that is predictable. The effect of the ionosphere
is frequency-dependent and very variable. The column density of
electrons is almost always below 100~TECU (one TECU corresponds to
$10^{16}~\text{m}^{-2}$), which at 1~GHz causes delays of about
130~nanoseconds. The delay scaling is inversely proportional to the
observing frequency squared and it may generally, thus, be neglected.

\subsection{Finding lensed FRBs}

The main caveat is that no gravitationally lensed FRB repeaters are
known thus far.  The ASKAP and CHIME telescopes are already finding
FRBs at a high rate, which, with a lensing fraction of about one in
1000, would mean about one lensed FRB per year. However, because they
cover only a small fraction of the sky at a given moment, they will
almost always miss the lensed copies of bursts. Researchers might try to
identify lensed candidates by their expected microlensing signature
and then try to find repeats and lensed copies with targeted
observations, but this approach would hardly find enough suitable
systems.

What is needed for the search of gravitationally lensed FRBs is a telescope
that continuously monitors a significant fraction of the sky without
interruption, so that no bursts or lensed images would be missed. A
good option is to concentrate on the circumpolar region, for example
north of the declination $50\degr$, which always has an elevation greater
than $10\degr$ at latitudes above $50\degr$. This patch corresponds to
more than one tenths of the entire sky, or about 20 times the field of
view of CHIME. At the same level of sensitivity as CHIME, we expect to find 40
FRBs per day \citep{fonseca20} or a substantial number of lensed ones
per year.

The most efficient instrument for such a wide-field monitoring is a
regular array of antennas or small dishes with a beamformer based on a
two-dimensional Fast Fourier Transform as proposed by \citet{otobe94}
and \citet{tegmark09}. A one-dimensional version of this approach is
already being used by CHIME-FRB \citep{ng17,masui19}, and an
application for the two-dimensional case is straight forward.
Appropriate antenna arrays were developed as design studies for the
Square Kilometre Array (SKA). The EMBRACE test array
\citep{torchinsky15,torchinsky16} is an $8\times 8~$m array consisting
of 4600 dual-polarisation (only one fully equipped) elements for the
frequency range between 900--1500~MHz. The original analogue beamformer cannot
directly be used for the FFT-based beamformer design, but we are
considering options to combine the existing array or a variant of it
with a CHIME-like beamformer as a first step in the direction of a
sensitive wide-field FRB search instrument.

\section{The gravitational lens as interferometer}
\label{sec:gravlens interferometer}

The notion of mutual coherence between gravitationally lensed images has already
been discussed by \citet{schneider85} and others. Within the framework of our discussion,
we can think of the gravitational lens as a huge intergalactic
interferometer, the arms of which are given by individual lensed
images. We assume that these have an apparent separation of
$\theta$, which corresponds to a baseline in the lens plane of
$L=\Dd\,\theta$. The angular resolution at a wavelength of
$\lambda\sub d$ (measured in the lens plane) is then $\lambda\sub
d/L$. The corresponding linear resolution in the source plane we call
$\Delta x$, which would be seen by the observer as the angular
resolution $\Delta \theta$, with
\begin{align}
  \Delta x &= \frac{\lambda_0}{\theta} \frac{\Dds}{(1+z\sub d)\,\Dd}
  , \\ \Delta \theta &= \frac{\lambda_0}{\theta\,\Deffcorr}
  .
\end{align}
Here, we assumed that the `arms of the interferometer' are fixed. In
reality, however, the images will slightly shift with position in the
source plane, which adds geometric and potential components to the phase
difference between parts of the source. Fortunately, these effects are cancelled out
because of Fermat's principle, as discussed in
Section~\ref{sec:non-exp lens}. A gravitational lens really is
equivalent to a fixed Young's double slit.

Using our order-of-magnitude estimates from Table~\ref{tab:numbers},
neglecting the lens redshift and the differences between distance
parameters, and using a wavelength of $\lambda_0=20~\text{cm}$, we get
a linear resolution of 40~km and an angular resolution of $3\times
10^{-16}~\text{arcsec}$, or about 0.3 femto-arcsec. With the actual
values from our lens sample, we find (for a 2~arcsec image separation)
linear and angular resolutions of 4--11~km and 15--53~atto-arcsec,
respectively.

The images are mutually coherent only as long as this
lens-interferometer does not resolve all their structures, in
other words, for only as long as the sources have structures on scales
below about ten kilometres. We emphasise that most of this paper
relies on this assumption because otherwise we could not achieve the
required precision in the measured redshifts.  \citet{cho20} find
narrow structures in FRB~181112 with rise time of down to 15~$\mu$s,
or 10~$\mu$s when correcting for redshift. This corresponds to a
maximum source size of 3~km, which is small enough to maintain
coherence.

As long as there is at least some level of mutual coherence, we can
now use the lens as interferometer with (for a quad) six independent
baselines. At the very least, we can compare amplitudes of correlations
(`visibilities') between baselines and in this way measure the sizes
of source components. It remains to be discussed how well the effects
of microlensing and interstellar scattering can be corrected for in
the analysis. These will be different for the different macro-images,
so that they can be treated as responses of the virtual interferometer
stations corresponding to the images. The resolution corresponding to
the subimages caused by these effects is much poorer, so that they
will not introduce differential effects across the FRB source. In
other words, we should be able to measure phase differences between the
visibilities of subcomponents of the source, or even between
subsequent bursts. Depending on the achievable S/N we may be able to
measure relative positions with accuracies of a kilometre or even
better. The extreme precision of such a measurement at cosmological
distances cannot be admired enough.

We cannot expect to measure consistent phases between bursts that are
separated more than the interstellar scintillation timescale.  With
the assumed reduced precision of a microsecond, we can measure
positional offsets between bursts with an accuracy on the order of
50~femto-arcsec or typically $10^4$~km, which is still impressive for
sources at Gpc distance. Uniform motion will be absorbed in the
relative proper motion, but any deviation from that, for example, due to
orbital motion, will be easily measurable with exquisite precision.
\citet{dai17} discuss this and other aspects of motion of the source,
lens, and observer in the context of lensed FRBs; however, these authors only
consider incoherent measurements with millisecond accuracy.

\section{Discussion}
\label{sec:discussion}

With gravitationally lensed FRBs, time delays can potentially be
measured very accurately, not only to milliseconds, as expected from
their short nature, but even to microseconds or even nanoseconds
because we can measure the wave fields of the images and correlate
them coherently. For repeating FRBs, this presents the opportunity to
measure how time delays change over timescales of months and years. These
changes are described as differential redshifts between the images. Here, we
derive how the differential redshifts are related to parameters of
the lens and the cosmological model. We find that in quadruply imaged
systems we have sufficient information to be able to combine the time delays
themselves with their time-derivatives (or the redshifts) to eliminate
uncertainties from the a priori unknown mass distribution of the lens
entirely.

In classical gravitational lensing, where mass models are fitted to
the observable image configuration (and perhaps other parameters such as
the velocity dispersion of the lens), fundamental parameter
degeneracies remain that cannot be removed without further
assumptions. Because the true mass distributions are unknown, it is
difficult to estimate the biases introduced by the assumptions. The
mass-model uncertainties are generally seen as the most serious source
of systematic errors in lensing.

Removing this main uncertainty with lensed FRBs does not come at no cost: the differential redshifts are extremely small and a number of
additional effects have to be taken into account in the analysis. Most
important is the relative transverse motion of the source, lens, and
observer. This contribution from proper motion is about three orders
of magnitude larger than the cosmological effect, but for quad systems
it can be determined and removed, as long as the relative image
positions are accurately known. Such measurements are possible with
VLBI.

\citet{kochanek96} argue that proper motion measurements at
cosmological distances would be very valuable to study our motion
relative to a rest frame defined by external galaxies independent of
the CMB dipole or to study local kinematics. They estimate that with
VLBI, normal peculiar velocities of galaxies are `too small for
direct detection given current life expectancies'. VLBI precision has
improved since then and life expectancies have generally increased, but
the statement is still true. The proper motion from
Table~\ref{tab:numbers} corresponds to less than one $\mu$arcsec over
a decade, which is beyond detectability with VLBI. \citet{kochanek96}
propose to measure quad lenses with VLBI to utilise the lensing
magnification to boost the motion and to provide nearby reference
sources but even then, obtaining a measurement is challenging. The relative
redshift induced by proper motion, on the other hand, enables very
accurate measurements and any conclusions will only be limited by the
sample variance. Large-scale systematic motion, such as cosmic
rotation on sub-horizon scales, can then be studied with high
accuracy, as well as other unexpected effects.

This is not the first time that time-derivatives of time delays have been
discussed in the literature. \citet{zitrin18} investigate a number of
aspects of observing cosmic expansion and other secular changes
(including transverse motion) in realtime. They calculate rates of
change of time delays for special mass models, but they do not combine time
delays and their derivatives to determine cosmological parameters, nor
do they describe how quad systems can be used to separate transverse
motion and cosmology.
\citet{piattella17} calculate expected drifts of redshifts, image
positions, and time delay for point-mass lenses, but do not discuss
which quantities can be measured with lensed FRBs.

At this moment it is not possible to estimate achievable accuracies
for cosmological parameters reliably because many important
parameters are unknown. Typical source redshifts are the most important in this regard,
but we also find that subtle details of the mass distribution of
lensing galaxies can lead to very different accuracies of our proposed
method. As a first attempt we used a simulated sample of lenses (not
claimed to be representative) as basis for an assessment of the
achievable results. With six lenses we find that accuracies of a few
percent may be realistic for the Hubble constant, the matter density,
and the equation-of-state parameter, $w,$ of the dark energy. We repeat
that these estimates are based on strong assumptions for the lenses
and, currently, for the assumed spatial flatness of the Universe.

Given that the classical approach of lensing time delays is producing
competitive results despite the known limitations, it should also be
applied to lensed FRBs.  \citet{li18} and \citet{liu19} argue that
lensed FRBs provide better constraints on the lens mass distribution
than lensed AGN because of the absence of a bright dominant component
in the host. Combining both methods is probably the best option. At
the very least, we can use classical modelling for lensed FRBs and use
the differential redshifts as additional constraints. Even if this
does not improve the cosmological result, using the time delays
themselves in addition to the redshifts provides one additional
constraint on the mass distribution -- it is only one because we add
three equations for the relative time delays but we also need to add the
source position as free parameter.

Further investigations of a number of aspects are needed, most
importantly: the effects of evolution of the mass distribution in the
main lens and along the line of sight. These can be estimated based on
structure formation simulations. Besides the unknown redshift
distribution, we may use lens models of existing well-studied non-FRB
lenses and determine how well those lenses would be suited for
cosmology based on the assumption that their sources could be FRBs.

Microlensing is another important effect. According to our estimates,
it does not invalidate our approach but, rather, it provides entirely new
information on the masses in the lenses and may in this way shed some
light on the dark matter problem. Time delays between micro-images
cannot be measured with other sources and their properties still have
to be investigated theoretically.

The physical nature of FRB sources is currently not understood at all,
but we know that they must either be small or at least have
subcomponents of at most a few kilometres in size. This means there is
no chance to resolve their structures with any astronomical
instrument, not even with space VLBI. Gravitational lenses can be used
as a natural telescopes, not just in the classical mode of  
magnifying the sources, but by providing interferometric baselines on the
scale of a galaxy. The electromagnetic fields of lensed images can be
correlated with each other and interferometric visibilities can be
produced, which would resolve structures of a few kilometres at
cosmological distances. This approach cannot be used for standard AGN
sources because they would be fully resolved out.  Having small
structures is not only necessary for measuring highly precise
coherent delays, but the correlations will resolve structures and help
studying the sources in turn.

Besides additional work on the theoretical understanding, it is now of
the highest importance that instruments are built that are equipped with the capacity to find
gravitationally lensed repeating FRBs in sufficient numbers. New
arrays such as ASKAP and CHIME completely changed the game regarding the
number of FRBs, but they are not optimised for the discovery of lensed
ones because they generally end up missing most of the lensed
images. Continuous coverage of a large field on the sky appears to be
the best way of finding lensed FRBs. The required technology exists,
along with even some of the hardware, and FFT telescopes can be built as efficient
lensed-FRB machines.
The potential for cosmology and other aspects of astrophysics is
certainly within reach and worth the effort.

\begin{acknowledgements}
LGS is a Lise Meitner independent research group leader and
acknowledges funding from the Max Planck Society.

Ue-Li Pen receives support from Ontario Research Fundresearch
Excellence Program (ORF-RE), Canadian Foundation for Innovation (CFI),
Simons Foundation, and Alexander von Humboldt Foundation. He
acknowledges the support of the Natural Sciences and Engineering
Research Council of Canada (NSERC, funding reference number
RGPIN-2019-067, CRD 523638-201).  He also receives substantial support
from the Max-Planck-Institut für Radioastronomie.
\end{acknowledgements}

\bibliographystyle{aa} \bibliography{lensed_frbs_published}

\onecolumn
\begin{appendix}
\section{Appendix: Details of the mock lens sample}

\begin{table*}[h]
  \caption{Lens model parameters: external shear, ellipticity of the
    potential, source position, scaling factor for time delays and the
    two coefficients for the redshifts. Lens and source redshifts are
    $z\sub d=0.5,0.9,0.4,0.6,0.3,0.7$, $z\sub
    s=0.8,1.6,0.7,1.0,0.6,1.5$.}
  \label{tab:lenses}
  \centering
  \begin{tabular}{lccccc@{}c}
    \hline\hline lens & $(\gamma_1,\gamma_2)$ & $\epsilon$ & $(x\sub
    s,y\sub s)~[\text{arcsec}]$ &
    $\Deffcorr/c~[\text{s}/\text{arcsec}^2]$ & $Z_t~[\kmsmpc]$ &
    $Z_\theta$ \\ \hline 1 & $(+0.033336,-0.084215)$ & $+0.071635$ &
    $(-0.124366,+0.000677)$ & $1.416725\times10^7$ & 8.930822 &
    $-1.192909$ \\ 2 & $(+0.062732,+0.081320)$ & $+0.049597$ &
    $(-0.109802,+0.071894)$ & $2.252066\times10^7$ & 8.818806 &
    $-1.899761$ \\ 3 & $(-0.014329,-0.046384)$ & $+0.009671$ &
    $(-0.026465,+0.009096)$ & $0.985719\times10^7$ & 8.291006 &
    $-0.838690$ \\ 4 & $(-0.033185,-0.012948)$ & $+0.052948$ &
    $(+0.028542,+0.009618)$ & $1.606895\times10^7$ & 9.239456 &
    $-1.346198$ \\ 5 & $(+0.005388,-0.040054)$ & $+0.001807$ &
    $(-0.028656,+0.015383)$ & $0.629996\times10^7$ & 7.225915 &
    $-0.545278$ \\ 6 & $(-0.063439,-0.099134)$ & $-0.005046$ &
    $(+0.093991,+0.043700)$ & $1.423638\times10^7$ & 9.288788 &
    $-1.191704$ \\ \hline
  \end{tabular}
\end{table*}

\begin{table*}[h]
  \caption{Parameters of all the lensed images: position, potential,
    time delay, and redshift due to Hubble expansion. The last three
    are relative to the first image. Images are labelled by increasing
    time delay.}
  \label{tab:images}
  \centering
  \begin{tabular}{cccccc}
    \hline\hline lens & image & $(x,y)~[\text{arcsec}]$ &
    $\psi~[\text{arcsec}^2]$ & $\Delta t~[10^6~\text{sec}]$ &
    $z~[10^{-11}]$ \\ \hline 1 & 1 & $(-0.616295,-1.045362)$ & 0 & 0 &
    0 \\ 1 & 2 & $(-0.104335,+1.102192)$ & $-0.128916$ & 0.958958 &
    $+0.373877$ \\ 1 & 3 & $(-0.946940,+0.292920)$ & $-0.429117$ &
    2.012286 & $+0.745334$ \\ 1 & 4 & $(+0.764052,-0.101282)$ &
    $-0.625325$ & 5.058688 & $+1.378106$ \\ 2 & 1 &
    $(-0.541751,+1.157300)$ & 0 & 0 & 0 \\ 2 & 2 &
    $(-0.113157,-1.030577)$ & $-0.293225$ & 4.923256 & $+1.386389$
    \\ 2 & 3 & $(-0.778740,-0.531192)$ & $-0.521266$ & 5.506729 &
    $+1.818991$ \\ 2 & 4 & $(+0.794084,-0.013921)$ & $-0.668530$ &
    8.971655 & $+2.493676$ \\ 3 & 1 & $(-0.882737,-0.595601)$ & 0 & 0
    & 0 \\ 3 & 2 & $(+0.428848,+0.936719)$ & $-0.038074$ & 0.222198 &
    $+0.077540$ \\ 3 & 3 & $(-0.612901,+0.762824)$ & $-0.131039$ &
    0.370791 & $+0.183847$ \\ 3 & 4 & $(+0.757856,-0.544997)$ &
    $-0.175040$ & 0.854613 & $+0.281791$ \\ 4 & 1 &
    $(+0.292660,+1.001393)$ & 0 & 0 & 0 \\ 4 & 2 &
    $(+0.127423,-1.004501)$ & $-0.032464$ & 0.399838 & $+0.111927$
    \\ 4 & 3 & $(+0.991349,-0.173856)$ & $-0.109075$ & 1.007788 &
    $+0.149528$ \\ 4 & 4 & $(-0.952484,+0.013629)$ & $-0.162293$ &
    1.877150 & $+0.342607$ \\ 5 & 1 & $(-0.900297,-0.540885)$ & 0 & 0
    & 0 \\ 5 & 2 & $(+0.261945,+1.000645)$ & $-0.018497$ & 0.072421 &
    $+0.022963$ \\ 5 & 3 & $(-0.694814,+0.707297)$ & $-0.095623$ &
    0.140376 & $+0.080222$ \\ 5 & 4 & $(+0.758091,-0.539219)$ &
    $-0.154889$ & 0.526483 & $+0.164609$ \\ 6 & 1 &
    $(+1.121534,+0.560593)$ & 0 & 0 & 0 \\ 6 & 2 &
    $(-0.899663,-0.479105)$ & $-0.266623$ & 3.351941 & $+0.847613$
    \\ 6 & 3 & $(-0.026164,-0.896564)$ & $-0.480632$ & 3.820875 &
    $+1.190042$ \\ 6 & 4 & $(-0.717676,+0.518437)$ & $-0.492224$ &
    3.883725 & $+1.220892$ \\ \hline
  \end{tabular}
\end{table*}

\end{appendix}

\end{document}